\documentclass[preprint,12pt]{elsarticle}
\usepackage{graphicx}
\usepackage{amssymb}
\usepackage{amsmath}
\usepackage{amsfonts} 
\usepackage{slashed} 
\usepackage[dvipsnames]{xcolor} 
\def\lamb#1#2{$^{#1}_{\Lambda}${#2}}

\newcommand{\be}{\begin{equation}} 
\newcommand{\ee}{\end{equation}}

\journal{Physics Letters B}

\begin{document}

\begin{frontmatter}

\title{Revisiting the hypertriton lifetime puzzle}  

\author[a]{A.~P\'{e}rez-Obiol}
\author[b]{~D.~Gazda} 
\author[c]{~E.~Friedman} 
\author[c]{~A.~Gal\corref{cor1}~}
\address[a]{Laboratory of Physics, Kochi University of Technology, 
Kami, Kochi 782-8502, Japan}
\address[b]{Nuclear Physics Institute, 25068 \v{R}e\v{z}, Czech Republic}
\address[c]{Racah Institute of Physics, The Hebrew University, Jerusalem 
91904, Israel}
\cortext[cor1]{corresponding author: Avraham Gal, avragal@savion.huji.ac.il} 

\date{\today} 

\begin{abstract} 

Conflicting values of the hypertriton (\lamb{3}{H}) lifetime were extracted in 
recent relativistic heavy-ion collision experiments. The ALICE Collaboration's 
reported \lamb{3}{H} lifetime $\tau$(\lamb{3}{H}) is compatible within 
measurement uncertainties with the free $\Lambda$ lifetime $\tau_\Lambda$, 
as naively expected for a loosely bound $\Lambda$ hyperon in \lamb{3}{H}, 
whereas STAR's reported range of $\tau$(\lamb{3}{H}) values is considerably 
shorter: $\tau_{\rm STAR}$(\lamb{3}{H})$\sim$(0.4-0.7)$\tau_{\Lambda}$. This 
\lamb{3}{H} lifetime puzzle is revisited theoretically, using \lamb{3}{H} 
three-body wavefunctions generated in a chiral effective field theory approach 
to calculate the decay rate $\Gamma$(\lamb{3}{H}$\,\to ^3$He$\,+\pi^-$). 
Significant but opposing contributions arise from $\Sigma NN$ admixtures 
in \lamb{3}{H} and from $\pi^-$-$^3$He final-state interaction. Evaluating 
the inclusive $\pi^-$ decay rate $\Gamma_{\pi^-}$(\lamb{3}{H}) via a branching 
ratio $\Gamma$(\lamb{3}{H}$\,\to ^{3}$He+$\pi^-)/\Gamma_{\pi^-}$(\lamb{3}{H}) 
determined in helium bubble-chamber experiments, and adding $\Gamma_{\pi^0}
$(\lamb{3}{H}) through the $\Delta I=\frac{1}{2}$ rule, we derive $\tau
$(\lamb{3}{H}) assuming several different values of the $\Lambda$ separation 
energy $B_{\Lambda}$(\lamb{3}{H}). It is concluded that each of ALICE and 
STAR reported $\tau$(\lamb{3}{H}) intervals implies its own constraint on 
$B_{\Lambda}$(\lamb{3}{H}): $B_{\Lambda}\lesssim 0.1$~MeV for ALICE, 
$B_{\Lambda}\gtrsim 0.2$~MeV for STAR. 

\end{abstract}


\end{frontmatter}

\section{Introduction}
\label{sec:intro}

The hypertriton, \lamb{3}{H}, a $\Lambda pn$ bound state with isospin $I$=0 
and spin-parity $J^P$=${\frac{1}{2}}^+$, is the lightest bound $\Lambda$ 
hypernucleus~\cite{GHM16}. Given the tiny $\Lambda$ separation energy 
$B_{\Lambda}$(\lamb{3}{H})=0.13$\pm$0.05~MeV~\cite{Davis05}, implying 
a $\Lambda$-deuteron mean distance of about 10~fm, the \lamb{3}{H} decay 
rate is expected to be close to that of the free $\Lambda$ hyperon which to 
99.7\% is governed by the nonleptonic $\Lambda$$\to$$N\pi$ weak-decay mode. 
This expectation was quantified using a $3N$ final-state closure approximation 
in early \lamb{3}{H} lifetime calculations~\cite{RD66} leading to an estimate 
$\Gamma$(\lamb{3}{H})/$\Gamma_\Lambda$=1+0.14$\sqrt{B_\Lambda}$ ($B_{\Lambda}$ 
in MeV), thereby suggesting a roughly 5\% enhanced \lamb{3}{H} decay 
rate $\Gamma$(\lamb{3}{H}) with respect to the free $\Lambda$ decay rate 
$\Gamma_\Lambda$, i.e. $\tau$(\lamb{3}{H})$\approx$0.95$\tau_{\Lambda}$. 
Yet values of $\tau$(\lamb{3}{H}) considerably shorter than $\tau_{\Lambda}$ 
were reported recently by two of the three relativistic heavy ion (RHI) 
experiments (HypHI and STAR) listed in Table~\ref{tab:tau}, in distinction 
from ALICE most recent values which within their own uncertainties agree 
with $\tau_{\Lambda}$~\cite{ALICE19,ALICE20}. In fact, a similarly large 
spread of $\tau$(\lamb{3}{H}) values, and with bigger uncertainties, had been 
reported in old nuclear emulsion and helium bubble-chamber (BC) hypernuclear 
measurements~\cite{Keyes73}. Finally, as if to compound confusion, STAR 
just published a $B_{\Lambda}$(\lamb{3}{H}) range of values much higher than 
listed above~\cite{STAR19,STAR20}. Implications to $A\leq 7$ hypernuclei 
are discussed in Ref.~\cite{Haiden20}. A more precise determination of 
$B_{\Lambda}$(\lamb{3}{H}) may constrain the balance between two-body 
hyperon-nucleon ($YN$) and three-body $YNN$ forces which at densities higher 
than met in hypernuclei is at the heart of the so called hyperon puzzle 
in neutron stars, i.e. the difficulty to reconcile recent observations 
of $\sim 2M_{\odot}$ neutron stars with having hyperons in their 
interior~\cite{LVB19,GKW20}; see Ref.~\cite{TF20} for a recent review. 

\begin{table}[htb] 
\begin{center} 
\caption{\lamb{3}{H} lifetime values $\tau$(\lamb{3}{H}) in ps from recent 
RHI experiments and from post-97 published, or to be published calculations. 
Note: $\tau_\Lambda$=263$\pm$2~ps~\cite{pdg}.} 
\begin{tabular}{ccc} 
\hline 
Exp/Th & Collaboration & $\tau$(\lamb{3}{H})  \\ 
\hline 
Exp & HypHI~\cite{HypHI13} & 183$^{+42}_{-32}\pm$37 \cite{HypHI13}  \\ 
Exp & STAR~\cite{STAR10,STAR18} & 142$^{+24}_{-21}\pm$29 \cite{STAR18}  \\ 
Exp & ALICE~\cite{ALICE16,ALICE19,ALICE20} & 242$^{+34}_{-38}\pm$17 
\cite{ALICE19}  \\  &  &  \\ 
Th & Kamada et al.~\cite{Kamada98} & 256  \\ 
Th & Gal-Garcilazo~\cite{GalGar19} & 213$\pm$5  \\ 
Th & Hildenbrand-Hammer~\cite{HH20} & $\approx \tau_{\Lambda}$  \\ 
\hline 
\end{tabular}  
\label{tab:tau} 
\end{center} 
\end{table} 

Table~\ref{tab:tau} also lists post-1997 $\tau$(\lamb{3}{H}) calculations 
known to us. The first calculation \cite{Kamada98} derived a value shorter by 
a few percent than $\tau_\Lambda$ in a complete Faddeev calculation, dealing 
with {\it all} three $\pi^-$ final-state channels: $^3$He$\,\pi^-$, $dp\pi^-$ 
and $ppn\pi^-$, while using the $\Delta I=\frac{1}{2}$ rule to add the $\pi^0$ 
decay channels. However, the $YN$ SC89 Nijmegen interaction \cite{NSC89} used 
there to construct a three-body \lamb{3}{H} wavefunction does poorly in 
hypernuclei, beginning with $A$=4 \cite{Nogga02}. The second calculation 
\cite{GalGar19} derived a $\tau$(\lamb{3}{H}) value shorter than 
$\tau_\Lambda$ by $\sim$20\%, half of which from attractive final-state 
interaction (FSI) of the outgoing pion. The third calculation \cite{HH20} 
is perhaps oversimplified by treating \lamb{3}{H} and $^3$He-$^3$H as 
$\Lambda d$ and $Nd$ loosely bound two-body systems, respectively. 

In this Letter we report on a new evaluation of the partial decay rate 
$\Gamma$(\lamb{3}{H}$\,\to ^{3}$He+$\pi^-$) using \lamb{3}{H} wavefunctions 
from a chiral effective field theory ($\chi$EFT) leading-order (LO) 
$YN$ interaction model~\cite{Polinder06,Haiden07} applied successfully 
in {\it ab initio} calculations of $A$=3,4 hypernuclear binding 
energies~\cite{Nogga13,Gazda14,Gazda16}. Surprisingly, the $\lesssim$0.5\% 
norm $\Sigma NN$ admixtures reduce by $\approx$10\% the purely $\Lambda NN$ 
decay rate. In contrast, using realistic low-energy $\pi^-{^3}$He distorted 
waves (DW) rather than plane waves (PW) enhances $\Gamma$(\lamb{3}{H}$\,
\to ^{3}$He+$\pi^-$) by $\approx$15\%. The inclusive $\pi^-$ decay 
rate $\Gamma_{\pi^-}$(\lamb{3}{H}) is then obtained using the BC world 
average branching ratio $R_3$=$\Gamma$(\lamb{3}{H}$\,\to ^{3}$He+$\pi^-)
/\Gamma_{\pi^-}$(\lamb{3}{H})=0.35$\pm$0.04 \cite{Keyes73}. Adding the 
inclusive $\pi^0$ decay rate, $\Gamma_{\pi^0}$(\lamb{3}{H})=$\frac{1}{2}
\Gamma_{\pi^-}$(\lamb{3}{H}) by the $\Delta I=\frac{1}{2}$ rule, we provide 
a new theoretical statement about $\tau$(\lamb{3}{H}) and its relationship 
to $B_{\Lambda}$(\lamb{3}{H}) for each of the three RHI experiments listed 
in Table~\ref{tab:tau}.

\section{Form Factors} 
\label{sec:FF} 

To relate $\Gamma$(\lamb{3}{H}$\,\to ^{3}$He+$\pi^-$) 
to the free-$\Lambda$ partial decay rate $\Gamma(\Lambda\to p\pi^-)$ 
we follow Kamada et al.~\cite{Kamada98}, Eq.~(A5), writing 
$\Gamma(\Lambda\to p\pi^-)$ in the form 
\begin{equation} 
\frac{\Gamma_{\Lambda\to p\pi^-}}{(G_Fm_{\pi}^2)^2}=\frac{k_{\pi^-}}{\pi}
\frac{M_p}{M_p+\omega_{\pi^-}}\left[{\cal A}_{\Lambda}^2+{\cal B}_{\Lambda}^2
\left(\frac{k_{\pi^-}}{2\bar{M}}\right)^2\right], 
\label{eq:Kamada} 
\end{equation}
with $k_{\pi^-}=100.6$, $\omega_{\pi^-}=172.0$, ${\bar M}=\frac{1}{2}
(M_p$+$M_\Lambda)=1027$, all in MeV, with $G_Fm_{\pi}^2=2.21\cdot 10^{-7}$, 
and where ${\cal A}_\Lambda=1.024$, ${\cal B}_\Lambda=-9.431$ are chosen 
here to satisfy the new BESIII value of the $\Lambda\to p\pi^-$ asymmetry 
parameter~\cite{BESIII}. This gives $\Gamma(\Lambda\to p\pi^-)=2.534$~GHz, 
and adding half of it for $\Gamma(\Lambda\to n\pi^0)$ to respect the 
$\Delta I=\frac{1}{2}$ rule yields $\tau_{\Lambda}\approx \tau(\Lambda
\to N\pi)$=$\Gamma^{-1}(\Lambda\to N\pi)$=263.1~ps. The squares 
of ${\cal A}_{\Lambda}$ and ${\cal B}_{\Lambda}$ above arise from 
a $\Lambda\to p\pi^-$ parity violating (PV) spin-independent amplitude 
${\cal A}_{\Lambda}$ and a parity conserving (PC) spin-dependent amplitude 
${\cal B}_{\Lambda}\vec{\sigma}\cdot{\hat{k}}_{\pi^-}$, respectively. The PV 
contribution in Eq.~(\ref{eq:Kamada}) dominates with 83\% of $\Gamma(\Lambda
\to p\pi^-)$. 

Proceeding to $\Gamma$(\lamb{3}{H}$\,\to ^{3}$He+$\pi^-$), we introduce 
nuclear form factors that accompany the $\Lambda\to p\pi^-$ PV and PC decay 
amplitudes, ${\cal A}_{\Lambda}\to {\cal A}_{\Lambda}\,F^{\rm PV}({\vec q})$ 
and ${\cal B}_{\Lambda}\vec{\sigma}\cdot{\hat{k}}_{\pi^-}\to{\cal B}_{\Lambda}
\,F^{\rm PC}({\vec q},\vec{\sigma})$: 
\begin{equation} 
\frac{\Gamma_{_{\Lambda}^3{\rm H}\to ^{3}{\rm He}+\pi^-}}{(G_Fm_{\pi}^2)^2}  
=3\frac{q}{\pi}\frac{M_{^3{\rm He}}}{M_{^3{\rm He}}+\omega_{\pi^-}(q)} 
\left[{\cal A}_{\Lambda}^2|F^{\rm PV}({\vec q})|^2+{\cal B}_{\Lambda}^2
|F^{\rm PC}({\vec q},\vec{\sigma})|^2\left(\frac{k_{\pi^-}}
{2\bar{M}}\right)^2\right] 
\label{eq:DL} 
\end{equation}
where the isospin factor 3 counts the three final nucleons to which the 
$\Lambda$ may turn into. Appropriate spin averages and summation are 
implied. In Eq.~(\ref{eq:DL}) the pion c.m. momentum (energy) is $q$=114.4 
($\omega_{\pi^-}$=179.3), $M_{^3{\rm He}}$=2809, using charge-averaged 
masses $M_N$=938.92, $m_\pi$=138.04, all in MeV. The nuclear form factors 
$F^j({\vec q},\vec{\sigma})$, where the index $j$ stands for PV or PC, 
are defined by 
\begin{equation} 
F^j({\vec q},\vec{\sigma})=\int{\Phi^{\ast}_f\,\phi_\pi({\vec q};{\vec r})\,
{\cal O}^j({\vec q},\vec{\sigma})\,\Phi_i\,{\rm d}^3r_3\,{\rm d}^3R}, 
\label{eq:Fj} 
\end{equation} 
where ${\cal O}^{\rm PV}=1$, ${\cal O}^{\rm PC}=\vec{\sigma}\cdot\hat{q}$, 
and $\Phi_{i,f}({\vec R},{\vec r_3})$ are initial \lamb{3}{H} and final 
$^3$He three-body wavefunctions in terms of Jacobi coordinates: $\vec R$ 
for the relative coordinate of spectator nucleons 1 and 2 and $\vec r_3$ 
for the coordinate of the third, `active' baryon relative to the c.m. of 
the spectator nucleons. Spin-isospin variables are kept implicit. 
The DW $\pi^-$ wavefunction $\phi_\pi({\vec q};{\vec r})$ evolves via FSI 
from a PW pion with momentum $\vec q$ in the \lamb{3}{H} rest frame. Its 
argument ${\vec r}=\frac{2}{3}{\vec r_3}$ is identified with the coordinate 
of the active baryon with respect to the c.m. of $^3$He. 

New $\Sigma$-hyperon two-body decay channels, $\Sigma^-$$\to$$n\pi^-$ and 
$\Sigma^0$$\to$$p\pi^-$, become available in \lamb{3}{H}$\,\to ^3$He+$\pi^-$ 
once $\Sigma NN$ admixtures are considered. The corresponding $\Sigma^-$ 
decay amplitudes are taken from studies of its weak decay: ${\cal A}_{
\Sigma^-}$=1.364, fitted to the lifetime value $\tau_{\Sigma^-}$=147.9$
\pm$1.1~ps~\cite{pdg}, and a negligible ${\cal B}_{\Sigma^-}$~\cite{DGH14}. 
Since the $\Sigma^0\to p\pi^-$ weak decay in free space is superseded by the 
$\Sigma^0\to\Lambda\gamma$ electromagnetic decay we use the chiral-Lagrangian 
prediction ${\cal A}_{\Sigma^0}$=$\frac{1}{\sqrt 2}{\cal A}_{\Sigma^-}
$~\cite{DGH14} and neglect ${\cal B}_{\Sigma^0}$. Using isospin basis 
consistent with that used in our \lamb{3}{H} wavefunction construction, the 
form factor $F^{\rm PV}$ in Eq.~(\ref{eq:DL}) is generalized according to: 
\begin{equation} 
{\cal A}_{\Lambda}F^{\rm PV}\to {\cal A}_{\Lambda}F_{I=0}^{\rm PV}+\frac{1}{3} 
({\sqrt 2}{\cal A}_{\Sigma^-}+{\cal A}_{\Sigma^0})F_{I=1}^{\rm PV}={\cal A}_{ 
\Lambda}F_{I=0}^{\rm PV}+\frac{1}{\sqrt 2}{\cal A}_{\Sigma^-}F_{I=1}^{\rm PV}, 
\label{eq:Sigma1} 
\end{equation} 
whereas ${\cal B}_{\Lambda}F^{\rm PC}\to {\cal B}_{\Lambda}F_{I=0}^{\rm PC}$. 
The subscripts $I=0,1$ indicate restricting in Eq.~(\ref{eq:Fj}) the 
expansion of the \lamb{3}{H} wavefunction $\Phi_i$ to $I_{NN}=0$ $\Lambda NN$ 
or to $I_{NN}=1$ $\Sigma NN$ components, respectively. Since the two PV 
amplitudes in Eq.~(\ref{eq:Sigma1}) interfere upon forming their summed 
absolute value squared, even as small a $\Sigma NN$ admixture probability as 
$P_{\Sigma} \lesssim 0.5\%$ may affect considerably the calculated \lamb{3}{H} 
two-body $\pi^-$ decay rate which is found to be {\it reduced} by slightly 
over 10\% from its value disregarding ${\cal A}_{\Sigma^-}$.

\section{Pion Distorted Waves} 
\label{sec:DW} 

The DW pion wavefunction $\phi_{\pi}({\vec q};{\vec r})$ input to the form 
factors $F^{\rm PV,PC}(\vec q)$, Eq.~(\ref{eq:Fj}), was generated from 
a standard optical potential~\cite{FG07,FG14}. The low-energy pion-nucleus 
interaction is well understood in terms of optical potentials constrained by 
pionic atoms data across the periodic table. Here we used optical potential 
parameters from large scale fits to $\pi^-$-atom level shifts and widths, 
from Ne to U~\cite{FG19,FG20}, where $s$- and $p$-wave $\pi N$ scattering 
amplitude parameters associated with optical potential terms linear in the 
nuclear density come out close to their threshold real on-shell values. 
Parameters associated with optical potential terms quadratic in density are 
phenomenological. Applying this potential to pionic atoms of $^3$He it is 
found to reproduce the experimental $1S$ level shift and width~\cite{3He84}. 

To extrapolate from near-threshold to $q$=114.4~MeV in the $\pi^-\,^3$He 
c.m. system we revised the above $\pi N$ linear-density terms using scattering 
amplitudes from the SAID package~\cite{SAID}. As for the non-linear terms, we 
extrapolated their threshold values by using also fits to $\pi^{\pm}$ elastic 
scattering at $T_{\rm lab}$=21.5~MeV on Si, Ca, Ni and Zr~\cite{PSI}. This 
resulted in a practically vanishing value of the $s$-wave term and a 65\% 
increase of the $p$-wave term. 

Expanding $\phi_{\pi}({\vec q};{\vec r})$ in our calculations in partial 
waves $\ell_{\pi}$, and recalling the spin-parity $J^P$=${\frac{1}{2}}^+$ 
of both $^3$He and \lamb{3}{H}, it follows that the only values allowed are 
$\ell_{\pi}$=0,2. 
Numerically we find a negligible $\ell_{\pi}$=2 contribution 
of order 0.1\%, proceeding exclusively through the relatively 
minor PC amplitude which in total contributes $\lesssim$3\% to 
$\Gamma$(\lamb{3}{H}$\,\to ^{3}$He+$\pi^-$). For the dominant 
$\ell_{\pi}$=0 contribution, $|F^{\rm PC}|^2=\frac{1}{9}|F^{\rm PV}|^2$
holds to better than 1\%. 


\begin{figure*}[!t] 
\begin{center} 
\includegraphics[width=0.7\textwidth]{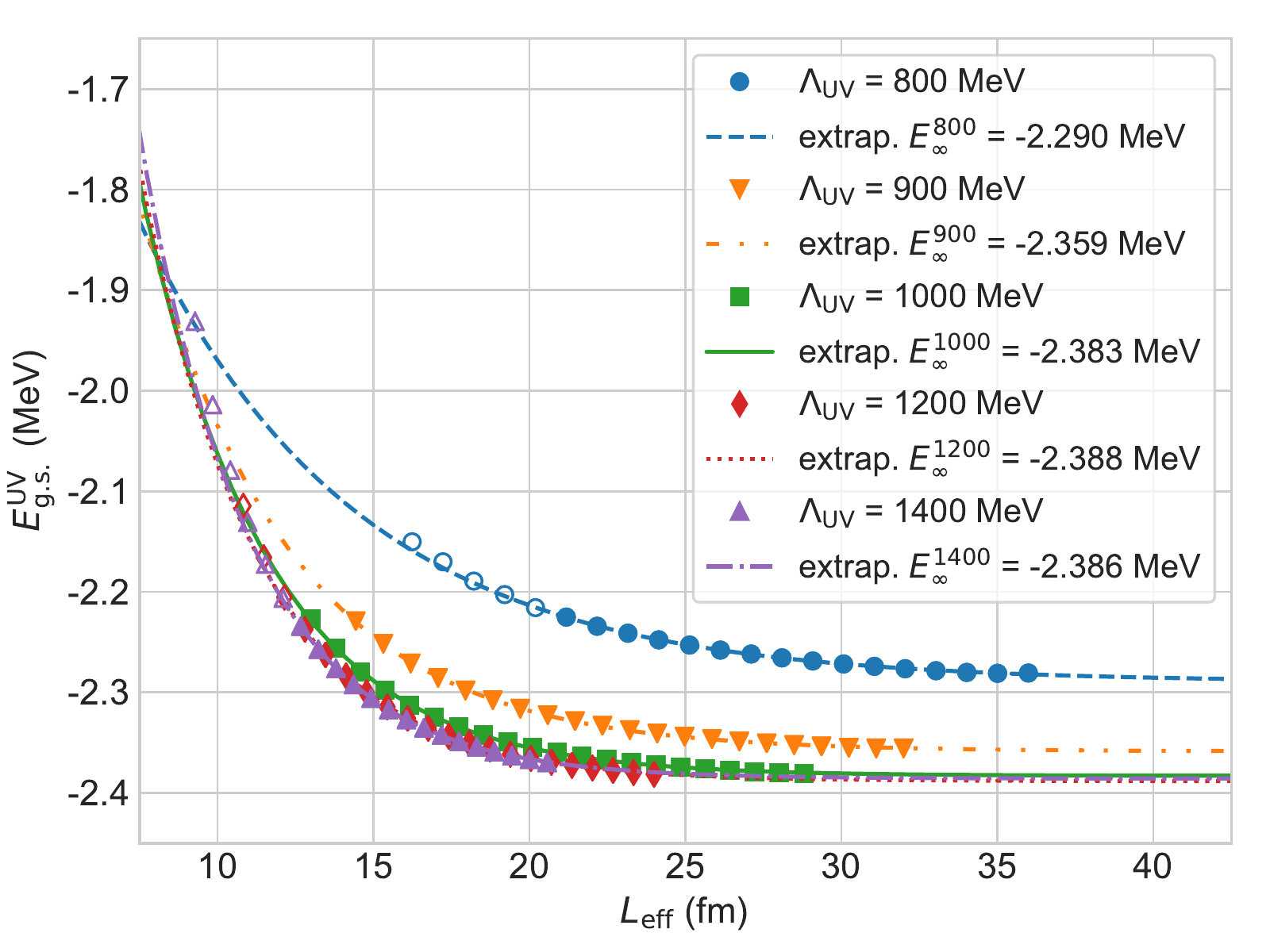} 
\includegraphics[width=0.7\textwidth]{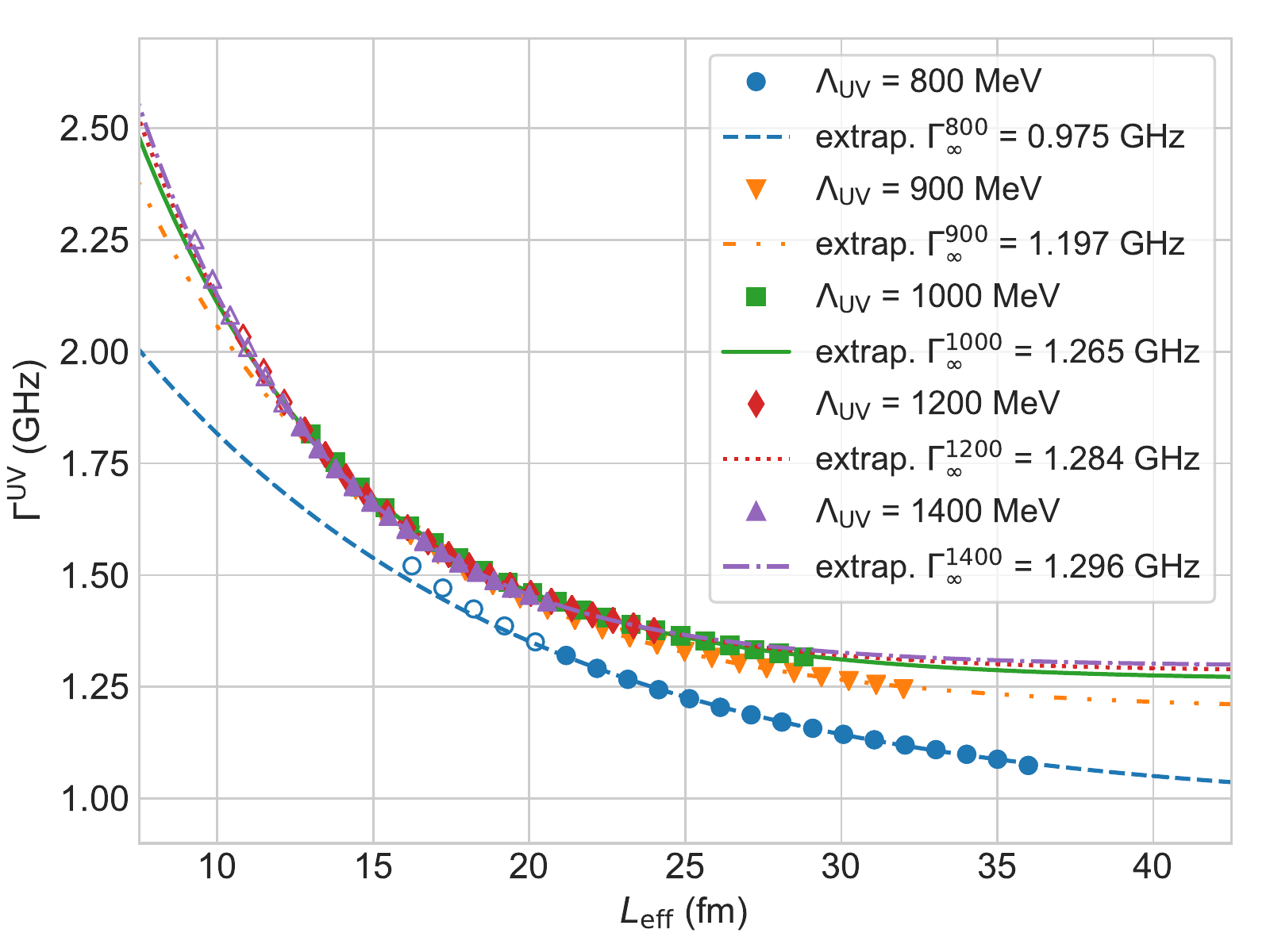} 
\caption{Extrapolations of NCSM calculated \lamb{3}{H} ground-state energies 
$E_{\rm g.s.}^{\rm UV}(L_{\rm eff})$ (upper) and the corresponding 
\lamb{3}{H}$\to ^3$He+$\pi^-$ decay rates $\Gamma^{\rm UV}(L_{\rm eff})$ 
(lower) as a function of the IR length $L_{\rm eff}$ using, e.g., 
Eq.~(\ref{eq:Leff}) for several fixed UV cutoff values $\Lambda_{\rm UV}$. 
$N_{\rm max}$=36 and $\hbar\omega$=14 MeV are held fixed for $^3$He. 
Open symbols start at $N_{\rm max}$=28. Filled symbols mark particle-stable 
\lamb{3}{H} configurations included in the fits, starting with a variable 
$N_{\rm max}$ between 28 and 40, and up to $N_{\rm max}$=68. $\Sigma NN$ 
admixtures and DW pions are included in the decay-rate calculations of the 
lower panel.} 
\label{fig:DW-LS} 
\end{center} 
\end{figure*}

\section{$YNN$ Input and $\Gamma$(\lamb{3}{H}$\,\to ^3$He+$\pi^-$)} 
\label{sec:input} 

Three-body wavefunctions of $^3$He and \lamb{3}{H}, input to $F^{\rm PV}$ and 
$F^{\rm PC}$ of Eq.~(\ref{eq:Fj}), were generated from Hamiltonians based on 
$\chi$EFT interactions: NNLO${\rm sim}$ for $NN+NNN$, 
derived by fitting $NN$ data up to $T_{\rm lab}^{\rm max}$=290 MeV with a 
regulator cutoff momentum $\Lambda_{\rm EFT}$=500 MeV~\cite{Carlsson16}, 
and LO $YN$ \cite{Polinder06,Haiden07} with $\Lambda_{\rm EFT}$=600 MeV. 
We followed the \emph{ab~initio} no-core shell model (NCSM) method within 
momentum-space harmonic-oscillator (HO) bases consisting of all excited states 
limited by ${\cal E}_{\rm HO}\leq (N_{\rm max}+3)\hbar\omega$ for a given HO 
frequency $\omega$~\cite{Wirth18}. The calculated $^3$He ground-state energy 
converges around $N_{\rm max}$=30 to $E_{\rm exp}(^3$He), independently of 
the HO frequency $\omega$ over a wide range. In contrast, convergence for the 
weakly bound \lamb{3}{H}, down to uncertainty of a few keV, is reached only 
in the largest NCSM space with $N_{\rm max}$=70. Although the \lamb{3}{H} 
energy computed at $N_{\rm max}$=70 exhibits a variational minimum for 
$\hbar\omega\approx 9$~MeV, with $B_{\Lambda}$(\lamb{3}{H})=0.16~MeV, 
the corresponding \lamb{3}{H}$\,\to ^3$He$\,+\pi^-$ decay rates exhibit 
undesired, stronger dependence on $\hbar\omega$. A standard empirical solution 
to tame such dependence is to extrapolate $\Gamma_{\hbar\omega}(N_{\rm max})$, 
for a fixed $\hbar\omega$, exponentially to $N_{\rm max}\to\infty$. 
Here, instead, we applied a recently proposed EFT-inspired extrapolation 
scheme, introducing infrared (IR) length scale $L_{\rm eff}$ and 
ultraviolet (UV) momentum scale $\Lambda_{\rm UV}$ to the NCSM many-body HO 
bases~\cite{Forssen18}. Fixing $\Lambda_{\rm UV}$ at a sufficiently large 
value, $\Lambda_{\rm UV}\gg\Lambda_{\rm EFT}$, the $\hbar\omega$ and $N_{\rm 
max}$ dependence of $\Gamma_{\hbar\omega}(N_{\rm max})$ may be traded off by 
its IR dependence on $L_{\rm eff}$. Extrapolating then $\Gamma_{\rm UV}(L_{
\rm eff})$, for a fixed $\Lambda_{\rm UV}$, exponentially to $L_{\rm eff}\to
\infty$, 
\begin{equation} 
\Gamma_{\rm UV}(L_{\rm eff}) = \Gamma^{\rm UV}_{\infty} + a^{\rm UV}
\exp(-2k^{\rm UV}L_{\rm eff}) 
\label{eq:Leff}  
\end{equation} 
with fit parameters $a^{\rm UV}, k^{\rm UV}$ and $\Gamma^{\rm UV}_{\infty}$, 
we obtained well-converged decay-rate values $\Gamma^{\rm UV}_{\infty}
$(\lamb{3}{H}$\,\to ^3$He+$\pi^-$), as shown in Fig.~\ref{fig:DW-LS} 
(lower panel) for several values of $\Lambda_{\rm UV}\geq 800$ MeV. This 
same procedure was applied, as shown in Fig.~\ref{fig:DW-LS} (upper panel), 
to extrapolate $E_{\rm g.s.}$(\lamb{3}{H})=$E_{\rm d}$$-$$B_{\Lambda}$, 
with $E_{\rm d}$ the calculated free deuteron energy. The figure exhibits UV 
convergence for $\Lambda_{\rm UV}\geq 1$~GeV, with rates $\Gamma^{\rm UV}_{
\infty}$(\lamb{3}{H}$\,\to^3$He+$\pi^-$)=1.28$\pm$0.02 GHz for $B_{\Lambda}^{
\rm UV}$(\lamb{3}{H})=0.162$\pm$0.003 MeV. The corresponding extraction 
uncertainties are estimated as 1~keV for $B_{\Lambda}$ and 3~MHz for $\Gamma$. 
Recalling the high-momentum cutoff scale $\Lambda_{\rm QCD}$$\sim$$M_B$, for 
a \lamb{3}{H} averaged baryon mass $M_B$$\approx$1~GeV, we chose to work with 
$\Lambda_{\rm UV}$=1~GeV. 

\begin{table}[htb] 
\begin{center} 
\caption{Partial decay rates $\Gamma^{\rm UV}_{\infty}$(\lamb{3}{H}$
\to ^3$He+$\pi^-$) obtained by adding up fixed $\Lambda_{\rm UV}$=1~GeV 
contributions from three leading configurations in $^3$He and in \lamb{3}{H} 
are listed in GHz for PW and DW pions. Each of these $l$=0 active-baryon 
($N,\Lambda,\Sigma$) configurations is specified by its spectator-nucleons 
isospin $I$, Pauli-spin $S$, orbital angular momentum $L$ and total angular 
momentum $J$. Probabilities $P$ are listed in percents. Total decay rates are 
given in the last row.} 
\begin{tabular}{cccccc} 
\hline 
$^{(2I+1)(2S+1)}L_J$ & $P$($^3$He) & $P_{\Lambda}$(\lamb{3}{H}) & 
$P_{\Sigma}$(\lamb{3}{H}) & $\Gamma^{\rm UV}_{\rm PW}$ & 
$\Gamma^{\rm UV}_{\rm DW}$ \\ 
\hline 
$^{13}S_1$ & 46.81 & 95.87 & -- & 1.141 & 1.310 \\ 
+$^{13}D_1$ & 48.42 & 99.23 & -- & 1.219 & 1.398 \\ 
+$^{31}S_0$ & 94.87 & 99.23 & 0.14 & 1.108 & 1.266 \\ 
\hline 
$L\leq 7$ $I,S\leq 1$ & 100 & 99.61 & 0.39 & 1.099 & 1.265 \\ 
\hline 
\end{tabular}  
\label{tab:wfs} 
\end{center} 
\end{table} 

The main contributions to $\Gamma^{\rm UV}_{\infty}$(\lamb{3}{H}$\,\to ^3$He+$
\pi^-$) at $\Lambda_{\rm UV}$=1~GeV, for both PW and DW pions, are listed 
in Table~\ref{tab:wfs}. As seen, just three leading $^3$He and \lamb{3}{H} 
configurations out of many other considered configurations reproduce within 
$\lesssim$1\% uncertainty the total \lamb{3}{H}$\to ^3$He+$\pi^-$ decay rate 
listed in the last row. Specifically, the dominant $P$(\lamb{3}{H})=96\% 
configuration in the first row corresponds to $s_{\Lambda}$ hyperon coupled 
to a $^3S_1$ quasi-deuteron that in $^3$He is close to the SU(4) limit of 
$P$=50\%. A few-percent $^3D_1$ $NN$ component induced by the tensor force 
in both $^3$He and \lamb{3}{H} is added in the second row, almost saturating 
$P$(\lamb{3}{H}). About half of the remaining 0.8\% probability arises from 
$\Sigma NN$ configurations, induced by the $\Lambda N\leftrightarrow\Sigma N$ 
transition component of the $\chi$EFT $YN$ LO potential~\cite{Polinder06}. 
The leading such configuration, listed in the third row, is $s_{\Sigma}$ 
hyperon coupled to a virtual-like $^1S_0$ $NN$ component that in $^3$He is 
again close to the SU(4) limit of $P$=50\%. Remarkably, this tiny $\Sigma NN$ 
admixture affects $\Gamma^{\rm UV}_{\infty}$(\lamb{3}{H}$\,\to ^3$He+$\pi^-$) 
more than the $NN$ tensor force does, reducing the two-body decay rate by 
$\approx$9\% as deduced by comparing the rates listed in the third row to 
those in the second row. The reduction is traced back to the sign of the 
$\Lambda N\leftrightarrow\Sigma N$ $^1S_0$ contact term in the $YN$ potential 
version used here; inverting this sign would reverse the sign of the observed 
charge symmetry breaking in the $A=4$ hypernuclei~\cite{Gazda16}. The use of 
DW pions increases the PW two-body decay rate by $\approx$15\%, inferred from 
the last row, higher than the $\approx$10\% found in Ref.~\cite{GalGar19} 
where the pion optical potential was limited to its $s$-wave part; the larger 
DW effect found here owes to including its $p$-wave part. The two effects 
recorded here work in opposite directions, combining to a merely 3\% increase 
of $\Gamma_{\rm DW}^{\Lambda + \Sigma}$(\lamb{3}{H}$\,\to ^{3}$He+$\pi^-$) 
with respect to $\Gamma_{\rm PW}^{\Lambda}$(\lamb{3}{H}$\,\to ^{3}$He+$\pi^-$) 
which is not listed here.

\section{From $\Gamma$(\lamb{3}{H}$\,\to ^3$He$\,+\pi^-$) to $\tau
$(\lamb{3}{H})} 
\label{sec:tau} 

To get the inclusive $\pi^-$ decay rate $\Gamma_{\pi^-}$(\lamb{3}{H}) from 
the two-body decay rate $\Gamma$(\lamb{3}{H}$\,\to ^{3}$He+$\pi^-$) we use 
the BC world average branching-ratio value $R_3=\Gamma$(\lamb{3}{H}$\,\to 
^{3}$He+$\pi^-)/\Gamma_{\pi^-}$(\lamb{3}{H})=0.35$\pm$0.04~\cite{Keyes73}. 
Note that decay tracks assigned to \lamb{3}{H} in BC experiments do not 
run the risk of resulting from decays of heavier hypernuclei in emulsion 
experiments where some of the tracks go unobserved and thereby potentially 
bias $B_{\Lambda}$ determinations, such as the world average value 
$B_{\Lambda}$(\lamb{3}{H})=0.13$\pm$0.05~MeV accepted by the hypernuclear 
community~\cite{Juric73}. 
Applying this $R_3$ to the two-body decay rate value 1.265~GHz associated 
in Fig.~\ref{fig:DW-LS} with $B_{\Lambda}$(\lamb{3}{H})=159~keV at 
$\Lambda_{\rm UV}$=1~GeV, and multiplying the obtained inclusive 
$\Gamma_{\pi^-}$(\lamb{3}{H}) by the $\Delta I=\frac{1}{2}$ factor 
$\frac{3}{2}$ so as to include $\Gamma_{\pi^0}$(\lamb{3}{H}), the resulting 
pionic-decay \lamb{3}{H} lifetime is $\tau_{\pi}(_{\Lambda}^3{\rm H})$=184$
\pm$21~ps, where the quoted uncertainty is statistical, arising from that of 
$R_3$. This calculated $\tau_{\pi}$(\lamb{3}{H}) is shorter by (30$\pm$8)\% 
than the free $\Lambda$ lifetime $\tau_\Lambda$=263$\pm$2~ps. The total 
lifetime $\tau$(\lamb{3}{H}) is shorter than that by (i) $\approx$1.5\% 
from $\Lambda N\to NN$ nonmesonic \lamb{3}{H} decay contributions 
\cite{RD66,Golak97,Axel18}; and (ii) $\approx$0.8\% from $\pi NN\to NN$ pion 
true absorption in \lamb{3}{H} decay (mostly two-body) channels, estimated 
within our pion optical potential. The $\approx$2.3\% summed yield of these 
non-pionic decay channels shortens slightly $\tau_{\pi}$(\lamb{3}{H}), leading 
to a \lamb{3}{H} lifetime $\tau(_{\Lambda}^3{\rm H})$=180$\pm$21~ps listed 
in the third row in Table~\ref{tab:sum}. It was tacitly assumed throughout 
this derivation of $\tau$(\lamb{3}{H}) that the branching ratio $R_3$ used 
here, taken from experiment~\cite{Keyes73}, indeed corresponds to $B_{\Lambda}
$(\lamb{3}{H})=159 keV at which the input $\Gamma$(\lamb{3}{H}$\,\to ^{3}$He+$
\pi^-$) was evaluated. 

\begin{table}[htb] 
\begin{center}  
\caption{Two-body decay rates $\Gamma$(\lamb{3}{H}$\,\to ^{3}$He+$\pi^-$) 
(GHz) calculated for several $\Lambda_{\rm UV}$ cutoffs (MeV) from \lamb{3}{H} 
wavefunctions at given $B_{\Lambda}$ values (keV), and as extrapolated to 
$B_{\Lambda}$=410 keV, along with lifetimes $\tau$(\lamb{3}{H}) (ps) evaluated 
using $R_3$=0.35$\pm$0.04~\cite{Keyes73}, the $\Delta I = \frac{1}{2}$ rule, 
and an added 2.3\% non-pionic decay rate.}  
\begin{tabular}{cccc} 
\hline 
$\Lambda_{\rm UV}$ & $B_{\Lambda}$ & $\Gamma$(\lamb{3}{H}$\,
\to ^{3}$He+$\pi^-$) & $\tau$(\lamb{3}{H}) \\ 
\hline 
800  & 69  & 0.975 & 234$\pm$27  \\ 
900  & 135 & 1.197 & 190$\pm$22  \\ 
1000 & 159 & 1.265 & 180$\pm$21  \\ 
 --  & 410 & 1.403 & 163$\pm$18  \\ 
\hline 
\end{tabular}  
\label{tab:sum} 
\end{center} 
\end{table}

\section{Relationship to $B_{\Lambda}$} 
\label{sec:B_L} 

Expecting a lifetime $\tau$(\lamb{3}{H}) close to $\tau_{\Lambda}$ for a 
weakly bound $\Lambda$ hyperon in \lamb{3}{H}, one might worry why the present 
fully microscopic UV-converged two-body rate calculation at $\Lambda_{\rm UV}
$=1~GeV yielded, when augmented by a branching ratio $R_3$ from experiment, 
a pionic lifetime $\tau_{\pi}$(\lamb{3}{H}) shorter than $\tau_{\Lambda}$ by 
as much as $\sim$30\%. In response we draw attention to the considerably lower 
two-body rates marked in Fig.~\ref{fig:DW-LS} for $\Lambda_{\rm UV}$=800, 
900~MeV, where UV convergence has not yet been fully achieved. This means 
that some UV corrections that depend on short-range details of the employed 
interactions are missing in the extrapolation scheme of Eq.~(\ref{eq:Leff}). 
Nevertheless, the correlation observed in the figure, for each value of 
$\Lambda_{\rm UV}$, between $\Gamma^{\rm UV}$(\lamb{3}{H}$\,\to ^{3}$He+$
\pi^-$) and its corresponding $B_{\Lambda}^{\rm UV}$(\lamb{3}{H}) appears 
robust. In particular the extrapolated two-body decay rates for $\Lambda_{
\rm UV}$=800, 900~MeV provide meaningfully converged rates using well 
converged \lamb{3}{H} wavefunctions with $B_{\Lambda}^{\rm UV}$=69, 135~keV 
respectively. Repeating for these two-body decay rates the procedure that led 
to a relatively short lifetime value in the third row of Table~\ref{tab:sum}, 
using the same $R_3$ value from experiment~\cite{Keyes73}, we obtain for 
the least bound \lamb{3}{H} case a value shorter than $\tau_{\Lambda}$ by 
only $\sim$11\%, as listed in first row of Table~\ref{tab:sum}. This value 
of $\tau$(\lamb{3}{H}) agrees reasonably with the latest published ALICE 
lifetime value~\cite{ALICE19} and, within its $R_3$ induced uncertainty, 
also with Kamada {\it et al.}'s lifetime value derived in a fully 
three-body calculation, both listed in Table~\ref{tab:tau}. Similarly, 
the lifetimes listed in the next two rows of Table~\ref{tab:sum} agree well 
within measurement uncertainties with the HypHI lifetime value listed in 
Table~\ref{tab:tau}. Hence, as long as all $B_{\Lambda}$ values within or 
close to the interval 0.07--0.16 MeV are acceptable, neither ALICE nor HypHI 
reported \lamb{3}{H} lifetime values may be excluded. Given that the 50~keV 
uncertainty in the cited value $B_{\Lambda}$(\lamb{3}{H})=0.13$\pm$0.05 MeV 
\cite{Davis05} is purely statistical, and that a systematic uncertainty of the 
same size is plausible, a conservative estimate of the combined uncertainty 
is 0.07 MeV, so that all values of $B_{\Lambda}$(\lamb{3}{H}) between 0.06 and 
0.20 MeV are acceptable, and so are both ALICE's and HypHI's lifetime values. 

To discuss more quantitatively STAR's reported $\tau$(\lamb{3}{H}) we 
extrapolate $\Gamma$(\lamb{3}{H}$\,\to ^{3}$He+$\pi^-$) from the calculated 
values listed in the first three rows of Table~\ref{tab:sum} to a decay-rate 
value appropriate to $B_{\Lambda}$(\lamb{3}{H})=0.41 MeV, STAR's mean value 
claimed recently: $B_{\Lambda}$(\lamb{3}{H})=0.41$\pm$0.12$\pm$0.11~MeV 
\cite{STAR20}. Expanding $\Gamma$(\lamb{3}{H}$\,\to ^{3}$He+$\pi^-$) 
in powers of $\sqrt{B_{\Lambda}}$, with just linear and quadratic terms, 
we derive a value $\Gamma$(\lamb{3}{H}$\,\to ^{3}$He+$\pi^-$) for 
$B_{\Lambda}$(\lamb{3}{H})=0.41 MeV, as listed in the last row of the table. 
Repeating the procedure explained above of obtaining $\tau$(\lamb{3}{H}), 
we get 163$\pm$18~ps which has substantial overlap with STAR's reported 
lifetime~\cite{STAR18} listed in Table~\ref{tab:tau}. In fact, had we used 
STAR's central value $R_3$=0.32 from their own observation of \lamb{3}{H} 
two-body and three-body $\pi^-$ decays, $R_3$=0.32$\pm$0.05$\pm$0.08 
\cite{STAR18}, we would have obtained $\tau$(\lamb{3}{H})=149~ps, almost 
coincident with STAR's central lifetime value listed in Table~\ref{tab:tau}.

\section{Concluding Remarks} 
\label{sec:concl} 

Reported in this work is a new microscopic 
three-body calculation of the \lamb{3}{H} pionic two-body decay rate $\Gamma
$(\lamb{3}{H}$\,\to ^{3}$He+$\pi^-$). Using the $\Delta I = \frac{1}{2}$ rule 
and a branching ratio $R_3$ from experiment to connect to additional pionic 
decay rates, the lifetime $\tau$(\lamb{3}{H}) was deduced. As emphasized here 
$\tau$(\lamb{3}{H}) varies strongly with the small, rather poorly known 
$\Lambda$ separation energy $B_{\Lambda}$(\lamb{3}{H}); it proves possible 
then to correlate each one of the three distinct RHI experimentally reported 
values $\tau_{\rm exp}$(\lamb{3}{H}) with a theoretical value $\tau_{\rm th}
$(\lamb{3}{H}) that corresponds to its own underlying $B_{\Lambda}
$(\lamb{3}{H}) value. The $B_{\Lambda}$(\lamb{3}{H}) intervals thereby 
correlated with these experiments are roughly $B_{\Lambda}\lesssim 0.1$~MeV, 
$0.1\lesssim B_{\Lambda}\lesssim 0.2$~MeV and $B_{\Lambda}\gtrsim 0.2$~MeV 
for ALICE, HypHI and STAR, respectively. New experiments proposed at 
MAMI on Li target~\cite{MAMI17} and at JLab, J-PARC and ELPH on $^3$He 
target~\cite{HYP18} will hopefully pin down precisely $B_{\Lambda}
$(\lamb{3}{H}) to better than perhaps 50~keV, thereby leading to 
a unique resolution of the `hypertriton lifetime puzzle'.   

\section*{Acknowledgments} 
We are grateful to Patrick Achenbach, Nir Barnea, Peter Braun-Munzinger, 
Benjamin D\"{o}nigus, Alessandro Feliciello, Hans-Werner Hammer, Ji\v{r}\'{i} 
Mare\v{s} and Satoshi Nakamura for useful remarks on a previous version. 
The work of DG was supported by the Czech Science Foundation, GA\v{C}R 
grant No. 19-19640S. Furthermore, the work of DG, EF and AG was partially 
funded by the European Union's Horizon 2020 research \& innovation 
programme, grant agreement 824093.

\end{document}